\documentclass{article}
\usepackage{arxiv}
\usepackage[utf8]{inputenc}
\usepackage{hyperref}
\usepackage{booktabs}
\usepackage{tikz}
\usetikzlibrary{matrix}
\usetikzlibrary{calc}
\usetikzlibrary{positioning}
\usetikzlibrary{decorations.pathreplacing}
\usepackage{amsmath}
\usepackage{array,multirow}
\usepackage{tcolorbox}
\usepackage{fontawesome}
\usepackage{graphicx}
\usepackage{grffile}
\usepackage{soul}
\usepackage{float}
\usepackage{placeins}

\title{TSSuBERT: Tweet Stream Summarization Using BERT}

\author{Alexis Dusart\\
    Universite de Toulouse UPS-IRIT\\
	118 route de Narbonne F- 31062 Toulouse cedex 9\\
	\texttt{alexis.dusart@irit.fr}\\
	\And
	Karen Pinel-Sauvagnat\\
	Universite de Toulouse UPS-IRIT\\
	118 route de Narbonne F- 31062 Toulouse cedex 9\\
	\texttt{karen.sauvagnat@irit.fr}\\
	\And
	Gilles Hubert\\
	Universite de Toulouse UPS-IRIT\\
	118 route de Narbonne F- 31062 Toulouse cedex 9\\
	\texttt{gilles.hubert@irit.fr}\\
}

\begin{document}

\maketitle

\begin{abstract}
    The development of deep neural networks and the emergence of pre-trained language models such as BERT allow to increase performance on many NLP tasks.
    However, these models do not meet the same popularity for tweet summarization, which can probably be explained by the lack of existing collections for training and evaluation.
    Our contribution in this paper is twofold : (1) we introduce a large dataset for Twitter event summarization, and (2) we propose a neural model to automatically summarize huge tweet streams. This extractive model combines in an original way pre-trained language models and vocabulary frequency-based representations to predict tweet salience. An additional advantage of the model is that it automatically adapts the size of the output summary according to the input tweet stream.
    We conducted experiments using two different Twitter collections, and promising results are observed in comparison with state-of-the-art baselines.
\end{abstract}

\definecolor{bluetweet}{RGB}{6,108,170}
\definecolor{redtweet}{RGB}{150,93,14}
\definecolor{verttweet}{RGB}{79,125,90}
\definecolor{yellowtweet}{RGB}{207,219,35}
\definecolor{orangetweet}{RGB}{219,169,31}
\definecolor{purpletweet}{RGB}{121,11,181}
\definecolor{browntweet}{RGB}{173,50,5}
\definecolor{greytweet}{RGB}{193,200,212}
\newcommand{\colorTwitterSignB}{\textcolor{bluetweet}{\faTwitter}}
\newcommand{\colorTwitterSignR}{\textcolor{redtweet}{\faTwitter}}
\newcommand{\colorTwitterSignV}{\textcolor{verttweet}{\faTwitter}}
\newcommand{\colorTwitterSignY}{\textcolor{yellowtweet}{\faTwitter}}
\newcommand{\colorTwitterSignO}{\textcolor{orangetweet}{\faTwitter}}
\newcommand{\colorTwitterSignP}{\textcolor{purpletweet}{\faTwitter}}
\newcommand{\colorTwitterSignM}{\textcolor{browntweet}{\faTwitter}}
\newcommand{\colorTwitterSignG}{\textcolor{greytweet}{\faTwitter}}

\newcommand{\colorTwitterSignBclair}{\textcolor{bluetweet!40}{\faTwitter}}
\newcommand{\colorTwitterSignVclair}{\textcolor{verttweet!40}{\faTwitter}}
\newcommand{\colorTwitterSignMclair}{\textcolor{browntweet!40}{\faTwitter}}
\newcommand{\colorTwitterSignGclair}{\textcolor{greytweet!40}{\faTwitter}}
\newcommand{\colorTwitterSignPclair}{\textcolor{purpletweet!40}{\faTwitter}}

\newcommand{\modelArchitecture}{
\begin{tikzpicture}
\node[text width=0.3cm] at (0,1.4) {50};
\draw[black,<->,line width=0.2mm] (-1,1.2)--(1,1.2);
\draw[black,rounded corners=5pt] (-1,0.7) rectangle (1,1.1);
\node[text width=1.4cm] at (0.3,0.9) {Tweet};
\draw[black,->,line width=0.2mm] (-0.8,0.7)--(-0.8,0.5);
\draw[black,->,line width=0.2mm] (0.8,0.7)--(0.8,0.5);
\draw[black,->,line width=0.2mm] (0,0.7)--(0,0.5);
\draw[black,rounded corners=5pt] (-1,-0.5) rectangle (1,0.5);
\node[text width=1.4cm] at (0,0) {DistilBERT};
\draw[black,->,line width=0.2mm] (-0.8,-0.5)--(-0.8,-0.7);
\draw[black,->,line width=0.2mm] (0.8,-0.5)--(0.8,-0.7);
\draw[black,->,line width=0.2mm] (0,-0.5)--(0,-0.7);
\node[text width=0.4cm] at (0,-0.9) {768};
\draw[black,<->,line width=0.2mm] (-1,-1.1)--(1,-1.1);
\draw[black,rounded corners=5pt] (-1,-1.6) rectangle (1,-1.2);
\node[text width=1.4cm] at (0,-1.4) {Embedding};

\node[text width=0.8cm] at (2.4,0.8) {30,525};
\draw[black,<->,line width=0.2mm] (1.2,0.6)--(3.8,0.6);
\draw[black,rounded corners=5pt] (1.2,0.1) rectangle (3.8,0.5);
\node[text width=2.4cm] at (2.5,0.3) {Tokens Frequency};
\draw[black,->,line width=0.2mm] (1.4,0.1)--(1.4,-0.1);
\draw[black,->,line width=0.2mm] (3,0.1)--(3,-0.1);
\draw[black,->,line width=0.2mm] (2.2,0.1)--(2.2,-0.1);
\draw[black,rounded corners=5pt] (1.2,-0.5) rectangle (3.2,-0.1);
\node[text width=1.8cm] at (2.3,-0.3) {Dense Layer};
\draw[black,->,line width=0.2mm] (1.4,-0.5)--(1.4,-0.7);
\draw[black,->,line width=0.2mm] (3,-0.5)--(3,-0.7);
\draw[black,->,line width=0.2mm] (2.2,-0.5)--(2.2,-0.7);
\node[text width=0.4cm] at (2.2,-0.9) {50};
\draw[black,<->,line width=0.2mm] (1.2,-1.1)--(3.2,-1.1);
\draw[black,rounded corners=5pt] (1.2,-1.6) rectangle (3.2,-1.2);
\node[text width=1.4cm] at (2.2,-1.4) {Embedding};

\draw[black,->,line width=0.2mm] (0.8,-1.6)--(0.8,-1.8);
\draw[black,->,line width=0.2mm] (1.4,-1.6)--(1.4,-1.8);
\draw[black,rounded corners=5pt] (0.1,-1.8) rectangle (2.1,-2.2);
\node[text width=1.8cm] at (1.2,-2) {Dense Layer};
\draw[black,->,line width=0.2mm] (0.3,-2.2)--(0.3,-2.4);
\draw[black,->,line width=0.2mm] (1.1,-2.2)--(1.1,-2.4);
\draw[black,->,line width=0.2mm] (1.9,-2.2)--(1.9,-2.4);

\node[text width=0.4cm] at (1.1,-2.6) {50};
\draw[black,<->,line width=0.2mm] (0.1,-2.8)--(2.1,-2.8);
\draw[black,rounded corners=5pt] (0.1,-2.9) rectangle (2.1,-3.5);
\node[text width=2cm] at (1.2,-3.1) {Dense Layer -};
\node[text width=1.4cm] at (1.1,-3.3) {Prediction};
\end{tikzpicture}
}

\newcommand{\modelArchitectureLarge}{
\begin{tikzpicture}
\draw[black,rounded corners=5pt] (-1,0.7) rectangle (1,1.1);
\node[text width=1.4cm] at (0.3,0.9) {Tweet};
\draw[black,->,line width=0.2mm] (-0.8,0.7)--(-0.8,0.5);
\draw[black,->,line width=0.2mm] (0.8,0.7)--(0.8,0.5);
\draw[black,->,line width=0.2mm] (0,0.7)--(0,0.5);
\draw[black,rounded corners=5pt] (-1,-0.7) rectangle (1,0.5);
\node[text width=1.5cm] at (0,-0.1) {Pre-trained Language Model};
\draw[black,->,line width=0.2mm] (-0.8,-0.7)--(-0.8,-0.9);
\draw[black,->,line width=0.2mm] (0.8,-0.7)--(0.8,-0.9);
\draw[black,->,line width=0.2mm] (0,-0.7)--(0,-0.9);

\draw[black,rounded corners=5pt] (-1,-1.3) rectangle (1,-0.9);
\node[text width=2cm] at (0.05,-1.1) {Representation};

\draw[black,rounded corners=5pt] (1.2,-0.7) rectangle (3.2,0.1);
\node[text width=2cm] at (2.5,-0.3) {Vocabulary Frequency};
\draw[black,->,line width=0.2mm] (1.4,-0.7)--(1.4,-0.9);
\draw[black,->,line width=0.2mm] (3,-0.7)--(3,-0.9);
\draw[black,->,line width=0.2mm] (2.2,-0.7)--(2.2,-0.9);

\draw[black,rounded corners=5pt] (1.2,-1.3) rectangle (3.2,-0.9);
\node[text width=2cm] at (2.25,-1.1) {Representation};

\draw[black,->,line width=0.2mm] (0.8,-1.3)--(0.8,-1.5);
\draw[black,->,line width=0.2mm] (1.4,-1.3)--(1.4,-1.5);

\draw[black,rounded corners=5pt] (0.1,-1.5) rectangle (2.1,-1.9);
\node[text width=1.4cm] at (1.1,-1.7) {Prediction};
\end{tikzpicture}
}

\newcommand{\problemOverview}{
\begin{tikzpicture}
\begin{scope}
\node[text width=1cm] at (-2,-0.25) {$t$};
\node[text width=1cm] at (0,0.5) {1};
\node[text width=1cm] at (2.3,0.5) {2};
\node[text width=1cm] at (4.6,0.5) {3};
\begin{scope}[xshift=1.6cm]
\node[text width=1cm] at (3.9,0.5) {4};
\node[text width=1cm] at (5.2,0.5) {5};
\node[text width=1cm] at (5.9,0.5) {6};
\node[text width=1cm] at (7.6,0.5) {7};
\end{scope}
\draw [decorate,decoration={brace,amplitude=5pt,raise=-3ex}]
  (-1,1) -- (4.8,1) node[midway]{Salience Prediction};
\draw [decorate,decoration={brace,amplitude=5pt,raise=-3ex}]
  (5,1) -- (9,1) node[midway]{Tweet Selection};

\node[text width=2cm] at (0,0) {\colorTwitterSignP\hspace{0.5em}\colorTwitterSignP};
% \node[text width=2cm] at (0,-0.25) {\colorTwitterSignB\hspace{0.5em}\colorTwitterSignB\hspace{0.5em}\colorTwitterSignB};
% \node[text width=2cm] at (0,-0.5) {\colorTwitterSignB\hspace{0.5em}\colorTwitterSignB};
\draw[black,->,line width=0.2mm] (0.5,-0.25)--(1,-0.25);
% \draw[black,-,line width=0.2mm] (1.2,-0.35)--(1.2,-0.15)--(2.6,-0.15)--(2.6,-0.35)--cycle;
\filldraw (1.3,0) circle (2pt);
\filldraw (1.3,-0.25) circle (2pt);
\filldraw (1.3,-0.5) circle (2pt);
\filldraw (1.8,-0.125) circle (2pt);
\filldraw (1.8,-0.375) circle (2pt);
\filldraw (2.3,-0.25) circle (2pt);
\draw[black,line width=0.2mm] (1.3,0) -- (1.8,-0.125);
\draw[black,line width=0.2mm] (1.3,0) -- (1.8,-0.375);
\draw[black,line width=0.2mm] (1.3,-0.25) -- (1.8,-0.125);
\draw[black,line width=0.2mm] (1.3,-0.25) -- (1.8,-0.375);
\draw[black,line width=0.2mm] (1.3,-0.5) -- (1.8,-0.125);
\draw[black,line width=0.2mm] (1.3,-0.5) -- (1.8,-0.375);
\draw[black,line width=0.2mm] (1.8,-0.125) -- (2.3,-0.25);
\draw[black,line width=0.2mm] (1.8,-0.375) -- (2.3,-0.25);
\draw[black,->,line width=0.2mm] (2.8,-0.25)--(3.3,-0.25);
\node[text width=2cm] at (4.6,0) {\colorTwitterSignPclair\hspace{0.5em}\colorTwitterSignPclair};
% \node[text width=2cm] at (4.6,-0.25) {\colorTwitterSignBclair\hspace{0.5em}\colorTwitterSignBclair\hspace{0.5em}\colorTwitterSignBclair};
% \node[text width=2cm] at (4.6,-0.5) {\colorTwitterSignBclair\hspace{0.5em}\colorTwitterSignBclair};
\node[black,text width=2cm] at (4.5,0) {\scriptsize 0.33\hspace{0.3em}0.07};
% \node[black,text width=2cm] at (4.5,-0.25) {\footnotesize 0.13\hspace{0.3em}0.11\hspace{0.3em}0.02};
% \node[black,text width=2cm] at (4.5,-0.5) {\footnotesize 0.52\hspace{0.3em}0.14};
\begin{scope}[xshift=1.6cm]
\draw[black,-,dashed,line width=0.2mm] (3.5,0)--(3.5,-0.5);
\node[text width=1cm] at (4.3,-0.25) {\colorTwitterSignP};
\draw[black,-,dashed,line width=0.2mm] (4.8,0)--(4.8,-0.5);
\node[text width=1cm] at (5.6,-0.25) {\colorTwitterSignP};
\draw[black,<->,line width=0.2mm] (6.1,-0.25)--(6.6,-0.25);
\draw[black,-,line width=0.2mm] (6.8,-0.5)--(6.8,0)--(7.5,0)--(7.5,-0.5)--cycle;
 \end{scope}
  
\begin{scope}[yshift=-1cm]
\node[text width=2cm] at (0,0) {\colorTwitterSignB\hspace{0.5em}\colorTwitterSignB\hspace{0.5em}\colorTwitterSignB};
\node[text width=2cm] at (0,-0.25) {\colorTwitterSignB\hspace{0.5em}\colorTwitterSignB\hspace{0.5em}\colorTwitterSignB};
\node[text width=2cm] at (0,-0.5) {\colorTwitterSignB\hspace{0.5em}\colorTwitterSignB};
\draw[black,->,line width=0.2mm] (0.5,-0.25)--(1,-0.25);
% \draw[black,-,line width=0.2mm] (1.2,-0.35)--(1.2,-0.15)--(2.6,-0.15)--(2.6,-0.35)--cycle;
\filldraw (1.3,0) circle (2pt);
\filldraw (1.3,-0.25) circle (2pt);
\filldraw (1.3,-0.5) circle (2pt);
\filldraw (1.8,-0.125) circle (2pt);
\filldraw (1.8,-0.375) circle (2pt);
\filldraw (2.3,-0.25) circle (2pt);
\draw[black,line width=0.2mm] (1.3,0) -- (1.8,-0.125);
\draw[black,line width=0.2mm] (1.3,0) -- (1.8,-0.375);
\draw[black,line width=0.2mm] (1.3,-0.25) -- (1.8,-0.125);
\draw[black,line width=0.2mm] (1.3,-0.25) -- (1.8,-0.375);
\draw[black,line width=0.2mm] (1.3,-0.5) -- (1.8,-0.125);
\draw[black,line width=0.2mm] (1.3,-0.5) -- (1.8,-0.375);
\draw[black,line width=0.2mm] (1.8,-0.125) -- (2.3,-0.25);
\draw[black,line width=0.2mm] (1.8,-0.375) -- (2.3,-0.25);
\draw[black,->,line width=0.2mm] (2.8,-0.25)--(3.3,-0.25);
\node[text width=2cm] at (4.6,0) {\colorTwitterSignBclair\hspace{0.5em}\colorTwitterSignBclair\hspace{0.5em}\colorTwitterSignBclair};
\node[text width=2cm] at (4.6,-0.25) {\colorTwitterSignBclair\hspace{0.5em}\colorTwitterSignBclair\hspace{0.5em}\colorTwitterSignBclair};
\node[text width=2cm] at (4.6,-0.5) {\colorTwitterSignBclair\hspace{0.5em}\colorTwitterSignBclair};
\node[black,text width=2cm] at (4.5,0) {\scriptsize 0.04\hspace{0.3em}0.48\hspace{0.3em}0.11};
\node[black,text width=2cm] at (4.5,-0.25) {\scriptsize 0.13\hspace{0.3em}0.11\hspace{0.3em}0.02};
\node[black,text width=2cm] at (4.5,-0.5) {\scriptsize 0.52\hspace{0.3em}0.14};
\begin{scope}[xshift=1.6cm]
\draw[black,-,dashed,line width=0.2mm] (3.5,0)--(3.5,-0.5);
\node[text width=1cm] at (4.3,-0.25) {\colorTwitterSignB\hspace{0.5em}\colorTwitterSignB};
\draw[black,-,dashed,line width=0.2mm] (4.8,0)--(4.8,-0.5);
\node[text width=1.5cm] at (5.6,-0.25) {\colorTwitterSignP\hspace{0.5em}\colorTwitterSignB\hspace{0.5em}\colorTwitterSignB};
\draw[black,<->,line width=0.2mm] (6.1,-0.25)--(6.6,-0.25);
\draw[black,-,line width=0.2mm] (6.8,-0.5)--(6.8,0)--(7.5,0)--(7.5,-0.5)--cycle;
\draw[bluetweet,-,line width=0.2mm] (7,-0.1)--(7.3,-0.1);
% \draw[black,-,line width=0.2mm] (7,-0.2)--(7.3,-0.2);
% \draw[black,-,line width=0.2mm] (7,-0.3)--(7.3,-0.3);
% \draw[black,-,line width=0.2mm] (7,-0.4)--(7.3,-0.4);
\end{scope}
% \node[text width=2cm] at (-0.7,1) {Tweets stream};
% \node[text width=3cm] at (2,1) {Relevance Prediction};
% \node[text width=3cm] at (3.5,0.5) {Relevance Threshold};
% \node[text width=3cm] at (5,1) {Similarity Threshold};
% \node[text width=2cm] at (6,0.5) {Time Window Summary};
% \node[text width=2cm] at (8,1) {Gold Standard};
\node[text width=1cm] at (-2,-0.25) {$t+1$};
\end{scope}
\end{scope}
\begin{scope}[yshift=-2cm]
\node[text width=2cm] at (0,0) {\colorTwitterSignG\hspace{0.5em}\colorTwitterSignG\hspace{0.5em}\colorTwitterSignG};
\node[text width=2cm] at (0,-0.25) {\colorTwitterSignG\hspace{0.5em}\colorTwitterSignG};
\node[text width=2cm] at (0,-0.5) {\colorTwitterSignG\hspace{0.5em}\colorTwitterSignG\hspace{0.5em}\colorTwitterSignG};
\draw[black,->,line width=0.2mm] (0.5,-0.25)--(1,-0.25);
% \draw[black,-,line width=0.2mm] (1.2,-0.35)--(1.2,-0.15)--(2.6,-0.15)--(2.6,-0.35)--cycle;
\filldraw (1.3,0) circle (2pt);
\filldraw (1.3,-0.25) circle (2pt);
\filldraw (1.3,-0.5) circle (2pt);
\filldraw (1.8,-0.125) circle (2pt);
\filldraw (1.8,-0.375) circle (2pt);
\filldraw (2.3,-0.25) circle (2pt);
\draw[black,line width=0.2mm] (1.3,0) -- (1.8,-0.125);
\draw[black,line width=0.2mm] (1.3,0) -- (1.8,-0.375);
\draw[black,line width=0.2mm] (1.3,-0.25) -- (1.8,-0.125);
\draw[black,line width=0.2mm] (1.3,-0.25) -- (1.8,-0.375);
\draw[black,line width=0.2mm] (1.3,-0.5) -- (1.8,-0.125);
\draw[black,line width=0.2mm] (1.3,-0.5) -- (1.8,-0.375);
\draw[black,line width=0.2mm] (1.8,-0.125) -- (2.3,-0.25);
\draw[black,line width=0.2mm] (1.8,-0.375) -- (2.3,-0.25);
\draw[black,->,line width=0.2mm] (2.8,-0.25)--(3.3,-0.25);
\node[text width=2cm] at (4.6,0) {\colorTwitterSignGclair\hspace{0.5em}\colorTwitterSignGclair\hspace{0.5em}\colorTwitterSignGclair};
\node[text width=2cm] at (4.6,-0.25) {\colorTwitterSignGclair\hspace{0.5em}\colorTwitterSignGclair};
\node[text width=2cm] at (4.6,-0.5) {\colorTwitterSignGclair\hspace{0.5em}\colorTwitterSignGclair\hspace{0.5em}\colorTwitterSignGclair};
\node[black,text width=2cm] at (4.5,0) {\scriptsize 0.60\hspace{0.3em}0.17\hspace{0.3em}0.09};
\node[black,text width=2cm] at (4.5,-0.25) {\scriptsize 0.10\hspace{0.3em}0.44};
\node[black,text width=2cm] at (4.5,-0.5) {\scriptsize 0.07\hspace{0.3em}0.12\hspace{0.3em}0.01};
\begin{scope}[xshift=1.6cm]
\draw[black,-,dashed,line width=0.2mm] (3.5,0)--(3.5,-0.5);
\node[text width=1cm] at (4.3,-0.25) {\colorTwitterSignG\hspace{0.5em}\colorTwitterSignG};
\draw[black,-,dashed,line width=0.2mm] (4.8,0)--(4.8,-0.5);
\node[text width=1.5cm] at (5.6,-0.25) {\colorTwitterSignP\hspace{0.5em}\colorTwitterSignB\hspace{0.5em}\colorTwitterSignB};
\node[text width=1cm] at (5.6,-0.5) {\colorTwitterSignG\hspace{0.5em}};
\draw[black,<->,line width=0.2mm] (6.1,-0.25)--(6.6,-0.25);
\draw[black,-,line width=0.2mm] (6.8,-0.5)--(6.8,0)--(7.5,0)--(7.5,-0.5)--cycle;
\draw[bluetweet,-,line width=0.2mm] (7,-0.1)--(7.3,-0.1);
\draw[greytweet,-,line width=0.2mm] (7,-0.2)--(7.3,-0.2);
\end{scope}
% \draw[black,-,line width=0.2mm] (7,-0.3)--(7.3,-0.3);
% \draw[black,-,line width=0.2mm] (7,-0.4)--(7.3,-0.4);
\node[text width=1cm] at (-2,-0.25) {$t+2$};
\end{scope}
\begin{scope}[yshift=-3cm]
\node[text width=2cm] at (0,0) {\colorTwitterSignV\hspace{0.5em}\colorTwitterSignV\hspace{0.5em}\colorTwitterSignV};
\node[text width=2cm] at (0,-0.25) {\colorTwitterSignV\hspace{0.5em}\colorTwitterSignV\hspace{0.5em}\colorTwitterSignV};
\node[text width=2cm] at (0,-0.5) {\colorTwitterSignV\hspace{0.5em}\colorTwitterSignV\hspace{0.5em}\colorTwitterSignV};
\draw[black,->,line width=0.2mm] (0.5,-0.25)--(1,-0.25);
% \draw[black,-,line width=0.2mm] (1.2,-0.35)--(1.2,-0.15)--(2.6,-0.15)--(2.6,-0.35)--cycle;
\filldraw (1.3,0) circle (2pt);
\filldraw (1.3,-0.25) circle (2pt);
\filldraw (1.3,-0.5) circle (2pt);
\filldraw (1.8,-0.125) circle (2pt);
\filldraw (1.8,-0.375) circle (2pt);
\filldraw (2.3,-0.25) circle (2pt);
\draw[black,line width=0.2mm] (1.3,0) -- (1.8,-0.125);
\draw[black,line width=0.2mm] (1.3,0) -- (1.8,-0.375);
\draw[black,line width=0.2mm] (1.3,-0.25) -- (1.8,-0.125);
\draw[black,line width=0.2mm] (1.3,-0.25) -- (1.8,-0.375);
\draw[black,line width=0.2mm] (1.3,-0.5) -- (1.8,-0.125);
\draw[black,line width=0.2mm] (1.3,-0.5) -- (1.8,-0.375);
\draw[black,line width=0.2mm] (1.8,-0.125) -- (2.3,-0.25);
\draw[black,line width=0.2mm] (1.8,-0.375) -- (2.3,-0.25);
\draw[black,->,line width=0.2mm] (2.8,-0.25)--(3.3,-0.25);
\node[text width=2cm] at (4.6,0) {\colorTwitterSignVclair\hspace{0.5em}\colorTwitterSignVclair\hspace{0.5em}\colorTwitterSignVclair};
\node[text width=2cm] at (4.6,-0.25) {\colorTwitterSignVclair\hspace{0.5em}\colorTwitterSignVclair\hspace{0.5em}\colorTwitterSignVclair};
\node[text width=2cm] at (4.6,-0.5) {\colorTwitterSignVclair\hspace{0.5em}\colorTwitterSignVclair\hspace{0.5em}\colorTwitterSignVclair};
\node[black,text width=2cm] at (4.5,0) {\scriptsize 0.13\hspace{0.3em}0.09\hspace{0.3em}0.05};
\node[black,text width=2cm] at (4.5,-0.25) {\scriptsize 0.13\hspace{0.3em}0.14\hspace{0.3em}0.01};
\node[black,text width=2cm] at (4.5,-0.5) {\scriptsize 0.18\hspace{0.3em}0.12\hspace{0.3em}0.02};
\begin{scope}[xshift=1.6cm]
\draw[black,-,dashed,line width=0.2mm] (3.5,0)--(3.5,-0.5);
% \node[text width=1cm] at (4.3,-0.25) {\colorTwitterSignB\hspace{0.5em}\colorTwitterSignB};
\draw[black,-,dashed,line width=0.2mm] (4.8,0)--(4.8,-0.5);
\node[text width=1.5cm] at (5.6,-0.25) {\colorTwitterSignP\hspace{0.5em}\colorTwitterSignB\hspace{0.5em}\colorTwitterSignB};
\node[text width=1cm] at (5.6,-0.5) {\colorTwitterSignG};
% \node[text width=1cm] at (5.6,-0.25) {\colorTwitterSignB\hspace{0.5em}};
\draw[black,<->,line width=0.2mm] (6.1,-0.25)--(6.6,-0.25);
\draw[black,-,line width=0.2mm] (6.8,-0.5)--(6.8,0)--(7.5,0)--(7.5,-0.5)--cycle;
\draw[bluetweet,-,line width=0.2mm] (7,-0.1)--(7.3,-0.1);
\draw[greytweet,-,line width=0.2mm] (7,-0.2)--(7.3,-0.2);
\draw[verttweet,-,line width=0.2mm] (7,-0.3)--(7.3,-0.3);
% \draw[black,-,line width=0.2mm] (7,-0.4)--(7.3,-0.4);
\end{scope}
\node[text width=1cm] at (-2,-0.25) {$t+3$};
\end{scope}
\begin{scope}[yshift=-4cm]
\node[text width=2cm] at (0,0) {\colorTwitterSignM\hspace{0.5em}\colorTwitterSignM\hspace{0.5em}\colorTwitterSignM};
\node[text width=2cm] at (0,-0.25) {\colorTwitterSignM\hspace{0.5em}\colorTwitterSignM};
% \node[text width=2cm] at (0,-0.5) {\colorTwitterSignB\hspace{0.5em}\colorTwitterSignB\hspace{0.5em}\colorTwitterSignB};
\draw[black,->,line width=0.2mm] (0.5,-0.25)--(1,-0.25);
% \draw[black,-,line width=0.2mm] (1.2,-0.35)--(1.2,-0.15)--(2.6,-0.15)--(2.6,-0.35)--cycle;
\filldraw (1.3,0) circle (2pt);
\filldraw (1.3,-0.25) circle (2pt);
\filldraw (1.3,-0.5) circle (2pt);
\filldraw (1.8,-0.125) circle (2pt);
\filldraw (1.8,-0.375) circle (2pt);
\filldraw (2.3,-0.25) circle (2pt);
\draw[black,line width=0.2mm] (1.3,0) -- (1.8,-0.125);
\draw[black,line width=0.2mm] (1.3,0) -- (1.8,-0.375);
\draw[black,line width=0.2mm] (1.3,-0.25) -- (1.8,-0.125);
\draw[black,line width=0.2mm] (1.3,-0.25) -- (1.8,-0.375);
\draw[black,line width=0.2mm] (1.3,-0.5) -- (1.8,-0.125);
\draw[black,line width=0.2mm] (1.3,-0.5) -- (1.8,-0.375);
\draw[black,line width=0.2mm] (1.8,-0.125) -- (2.3,-0.25);
\draw[black,line width=0.2mm] (1.8,-0.375) -- (2.3,-0.25);
\draw[black,->,line width=0.2mm] (2.8,-0.25)--(3.3,-0.25);
\node[text width=2cm] at (4.6,0) {\colorTwitterSignMclair\hspace{0.5em}\colorTwitterSignMclair\hspace{0.5em}\colorTwitterSignMclair};
\node[text width=2cm] at (4.6,-0.25) {\colorTwitterSignMclair\hspace{0.5em}\colorTwitterSignMclair};
\node[black,text width=2cm] at (4.5,0) {\scriptsize 0.04\hspace{0.3em}0.10\hspace{0.3em}0.05};
\node[black,text width=2cm] at (4.5,-0.25) {\scriptsize 0.50\hspace{0.3em}0.08};
\begin{scope}[xshift=1.6cm]
\draw[black,-,dashed,line width=0.2mm] (3.5,0)--(3.5,-0.5);
\node[text width=1cm] at (4.3,-0.25) {\colorTwitterSignM};
\draw[black,-,dashed,line width=0.2mm] (4.8,0)--(4.8,-0.5);
\node[text width=1.5cm] at (5.6,-0.25) {\colorTwitterSignP\hspace{0.5em}\colorTwitterSignB\hspace{0.5em}\colorTwitterSignB};
\node[text width=1cm] at (5.6,-0.5) {\colorTwitterSignG\hspace{0.5em}\colorTwitterSignM};
\draw[black,<->,line width=0.2mm] (6.1,-0.25)--(6.6,-0.25);
\draw[black,-,line width=0.2mm] (6.8,-0.5)--(6.8,0)--(7.5,0)--(7.5,-0.5)--cycle;
\draw[bluetweet,-,line width=0.2mm] (7,-0.1)--(7.3,-0.1);
\draw[greytweet,-,line width=0.2mm] (7,-0.2)--(7.3,-0.2);
\draw[verttweet,-,line width=0.2mm] (7,-0.3)--(7.3,-0.3);
% \draw[browntweet,-,line width=0.2mm] (7,-0.4)--(7.3,-0.4);
\end{scope}
\node[text width=1cm] at (-2,-0.25) {$t+4$};
\end{scope}
\end{tikzpicture}
}
\section{Introduction}

The emergence of social media such as Twitter offers to users a considerable amount of data and information, that are sometimes not covered by traditional media.
This new and fresh information may lead to an information overload that could be lighten by automatic summarization. Although many approaches can be found in the literature \cite{DBLP:conf/sigir/RudraGGM018,DBLP:journals/tcss/RudraGGIM19, DBLP:conf/eacl/Olariu14, DBLP:journals/taslp/ZhangLGY13}, automatic summarization of huge amount of data is still an open issue \cite{ELKASSAS2021113679}.

At the same time, the recent state of the art in automatic summarization is mainly based on pre-trained language models \cite{DBLP:conf/emnlp/LiuL19} such as BERT \cite{DBLP:conf/naacl/DevlinCLT19}, following the example of the majority of NLP tasks. To the best of our knowledge such pre-trained language models have not been used for stream summarization, except the very recent work in \cite{li2021twitter} that focuses on specific summaries answering who-what-when-where questions. This scarcity might be due to the lack of datasets large enough to learn neural network models.

In this context, we first propose in this paper an incremental approach for Twitter stream summarization.  This approach has two original aspects: i) it combines BERT pre-trained model with tweet context to predict tweet salience, and (ii) it automatically estimates the appropriate size of the summary to propose at a given time.
Second, to overcome the lack of training data we introduce a new large and publicly available dataset for event summarization, named TES 2012-2016.
We conducted several experiments on this dataset and another smallest one used by state-of-the-art approaches. The observed results showed the benefit of our approach and unveil new open issues for the community.

\section{Related Work}
\label{sec:rw}

 Recent document summarization approaches are all based on neural architectures  used to generate extractive \cite{DBLP:conf/acl/ZhongLCWQH20} or  abstractive  summaries \cite{DBLP:conf/icml/ZhangZSL20, DBLP:journals/corr/abs-2010-08014, DBLP:conf/emnlp/QiYGLDCZ020}, and even both of them \cite{DBLP:conf/aaai/SongWFL020,DBLP:conf/emnlp/LiuL19}. Pre-trained language models such as BERT and BART are used in most of them and now serve as baselines on reference datasets such as CNN/DailyMail \cite{DBLP:conf/acl/SeeLM17,Hermann:2015}, Gigaword \cite{graff2003english,Rush_2015}, and X-Sum \cite{DBLP:conf/emnlp/NarayanCL18}.
 
 Tweet summarization, as a special case of multi-document summarization, raises particular issues linked to the volume of data and stream processing \cite{DBLP:journals/jips/RudrapalDB18}. TREC  microblog between 2011 and 2015 \cite{lin2014overview}, TREC RTS (Real Time Summarization) between 2016 and 2018 \cite{DBLP:conf/trec/SequieraTL18}, and TREC IS (Incident Stream) since \cite{mccreadie2020incident}, provided reference test collections to the community. Although many extractive summarization approaches have been proposed in these frameworks, the considered research scenarios remain very specific (e.g., monitoring channels for emergency operators for TREC IS) or have some biases \cite{hubert:hal-02319737}.
 
Among the most recent approaches for tweet summarization, extractive approaches named COWTS \cite{DBLP:conf/cikm/RudraGGGG15}, SEMCOWTS \cite{DBLP:journals/tweb/RudraGGG18}, and SCC \cite{DBLP:conf/sigir/RudraGGM018}  are based on Integer Linear Programming (ILP) to maximize
the coverage of content words in the final summary.
Authors more recently extended their approach to abstractive summarization \cite{DBLP:journals/tcss/RudraGGIM19}.
The use of pre-trained language model is mentioned in \cite{DBLP:conf/www/LiZ20} for abstractive summarization and very recently in \cite{li2021twitter} for extractive summarization. The latter uses BERT to encode tweets as well as a tweet relation graph as input of a graph convolutional network to generate tweet hidden features for tweet salience estimation. A non-publicly available collection was used to evaluate summaries answering who-what-where-when questions.

      In this paper, we present an extractive multi-document summarization method for Twitter stream summarization. Inspired by the method of \cite{DBLP:conf/emnlp/LiuL19} for text summarization, our model takes advantage of pre-trained language models and neural networks. 
Regarding the tweet stream context, our method generates summaries in an incremental way, which differs from the above presented methods. It also includes an original way to take into account tweet context. 
\section{TSSuBERT: a Tweet Stream Summarization approach Using BERT}
\label{sec:model}

\subsection{Problem Formulation}
\label{subsec:problem_formulation}

We formalize the issue as follows.
Given an event $e$, a tweet stream $T_{e,t}$ concerning $e$ between 2 timestamps $t_0$ and $t$, our aim is to build a summary $S_{e,t_0,t}$ of $T_{e,t}$,  $S_{e,t_0,t}$  being an ordered subset of $T_{e,t}$.

As $T_{e,t}$ may represent a huge volume of information (several million of tweets), it does not seem reasonable to start $T_{e,t+1}$ from scratch, i.e., by considering the whole stream  $T_{e,t+1}$. 
As a consequence, the problem can be considered as an incremental one as follows: $S_{e,t_0,t+1} = S_{e,t_0,t} \cup S_{e,t, t+1}$

\subsection{Architecture}

\begin{figure*}[!ht]
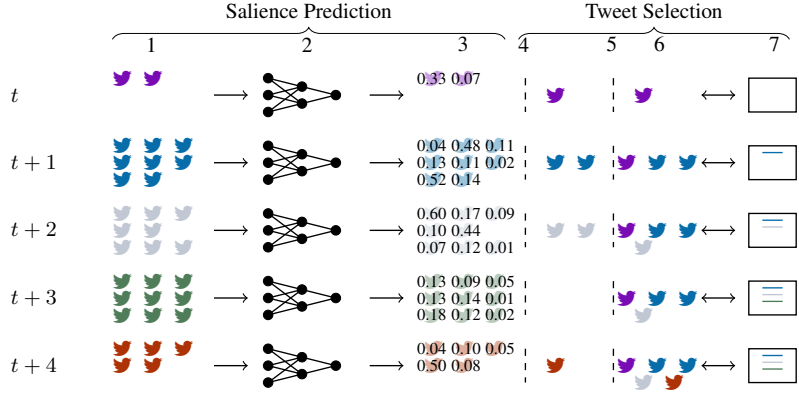

    \centering
    \scalebox{0.9}{
    \small
    \problemOverview
    }
    \caption{\label{fig:overview}\small{Overview of the model architecture for an event with 5 time increments ($t$ to $t+4$). Each column corresponds to: (1) the tweet stream for the considered time increment, (2) the salience score prediction, (3) the output of the model, i.e. tweets with their salience score, (4) tweets filtering according salience, (5) tweets filtering according to similarity w.r.t. the existing summary, (6) the incremental extractive result summary, and (7) the gold standard up to this time increment.}}
\end{figure*}

We propose a 2-phase architecture to tackle the problem (see Figure \ref{fig:overview}): (1) as summarizing a whole stream of tweets in input of a neural model have no sense, our model first predicts the salience of a tweet with a pre-trained language model combined with vocabulary frequencies, (2) tweets being above a given relevance threshold are then selected according to their similarity with the existing summary.

\begin{figure}[!ht]
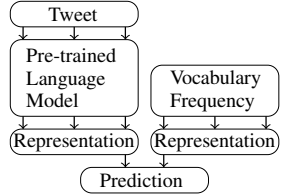

    \centering
    \scalebox{0.85}{
    \small
    \modelArchitectureLarge
    }
    \caption{\label{fig:overallarchitecture}Architecture of the salience prediction part.}
\end{figure}

\paragraph{Salience prediction}
The architecture of the salience prediction part of our model is presented in Figure \ref{fig:overallarchitecture}.
The aim of this part is to decide if a tweet should be kept in the event summary. A first and naive idea  would be to predict a positive or negative label regarding each tweet. This is however not conceivable, given the very unbalanced nature of the stream. Indeed, million of tweets can be written regarding an event, although only a few of them (fewer than tens of them) will be kept for the summary.

As a consequence, we decided to predict a salience score for each tweet. For this purpose, we take advantage of state-of-the-art text representations, i.e., pre-trained language models such as BERT \cite{DBLP:conf/naacl/DevlinCLT19}. 
Such representations have generally been applied to specific natural language understanding tasks by fine-tuning a final output layer for the task. 

In addition, predict the salience of a tweet without any contextual information on the event would not allow to generalize the model to other events.
We thus decided to integrate the context of the tweets using the whole vocabulary related to the event, i.e., the frequencies of the terms appearing in the event stream. 

Two contributions can be highlighted from this simple architecture: 
(i) the combined use of a pre-trained language model and vocabulary frequencies representations to predict salience;
(ii) the use of a pre-trained language model in itself for extractive tweet summarization, which, surprisingly, cannot be found in literature  (except the work in \cite{li2021twitter}).

\paragraph{Tweet selection}

The tweets with a salience score above a given threshold $\lambda_{salience}$ are then considered for inclusion in the summary.
To do so, we introduce a similarity score $sim(c, S)$  between each candidate tweet $c$ and the existing summary $S$. 
For each tweet in the existing summary, we compute its similarity with the candidate tweet. If the similarity score is lower than a $\lambda_{similarity}$ threshold for all the tweets in the existing summary, the candidate tweet is kept for the summary.

A strength of our model lies in this tweet selection based on two thresholds: unlike most state-of-the-art models we do not need to give a fixed summary size as input of the model, since it automatically adapts the size of the output summary according to the input tweet stream.

\section{Experiments and results}
\label{sec:exp}

\subsection{A new collection for tweet summarization: TES 2012-2016}

To the best of our knowledge, tweet summarization collections large enough to train neural models are not available to the community.
We thus built such a collection using an existing tweet event dataset. 
We retrieved tweets from the set of ids provided in the Twitter events 2012-2016 collection  \cite{DBLP:journals/jasis/Zubiaga18}. This dataset is composed of around 150 million of tweet ids crawled for 30 events between 2012 to 2016. Tweets were crawled according to different sets of keywords related to the considered events.
At the time we did the experiments, we could retrieve 82 million of tweets.
We built Gold Standard following the methodology in \cite{DBLP:conf/acl/GhalandariHPGI20} by using the Wikipedia Current Event portal\footnote{\url{https://en.wikipedia.org/wiki/Portal:Current_events}}.
For each event, we extracted the existing summaries in the portal pages for the concerned period. As summaries are presented in a daily manner in the Wiki Portal, the time increment for each summary is fixed to one day.
We finally considered 28 events out of 30, since 2 events were not listed in the Wikipedia Current Event portal.

In addition, we constructed Oracle summaries for the collection, in order to know the reachable upper bounds. Two different Oracle summaries were built, 
according to the two different relevance metrics ROUGE-2 and Cosine. To do so, for each day of the event we used a greedy algorithm that keeps the tweets maximising the metric, as done in \cite{DBLP:conf/emnlp/LiuL19,DBLP:conf/aaai/NallapatiZZ17,DBLP:journals/corr/abs-2010-08014}. 

Some statistics about the collection are available in Table \ref{tab:stats_descs_collections}, line 2. The TES 2012-2016 collection is available on our (anonymous) Github\footnote{\url{https://github.com/JoeBloggsIR/TSSuBERT}}. To the best of our knowledge, this is the biggest collection available for tweet summarization.

\begin{table}[t]
\centering
\scalebox{0.85}{
\begin{tabular}{p{4.5cm}cccc}
    \toprule
    \multirow{2}{*}{Collection} & \multirow{2}{*}{\# streams} & \# days & Mean & Mean Gold \\
    & & (total) & \# tweets/event & Standard length\\
    \midrule
    TES 2012-2016 & 28 & 545 & 2,939,842 & 265 (22)\\
    \cite{DBLP:conf/sigir/RudraGGM018} collection & 11 & 30 & 7,261 & 540 (198)\\
    \bottomrule
\end{tabular}
}

\caption{\label{tab:stats_descs_collections}\small{Some statistics for the TES 2012-2016 collection and the collection in \cite{DBLP:conf/sigir/RudraGGM018}. The number in () for the gold standard length represents the average number of words added at each time increment, i.e., each day.}}
\end{table}
\normalsize

\subsection{Experiments on TES 2012-2016}

\begin{table*}[!htbp]
\centering
\scalebox{0.75}{
\begin{tabular}{cp{2.2cm}ccccccc}
    \toprule
    &\multirow{2}{*}{Model} & \multicolumn{2}{c}{ROUGE-1 F} & \multicolumn{2}{c}{ROUGE-2 F} & \multicolumn{2}{c}{COS Embed} & Max Length per\\
    && micro & macro & micro & macro & micro & macro & time increment, i.e., per day\\
    \midrule
     \multirow{7}{*} {TES 2012-2016} &Oracle R-2 & 0.532** & 0.554** & 0.330** & 0.357** & 0.907** & 0.903** & G.S. length\\
    &Oracle COS & 0.575** & 0.591** & 0.218** & 0.245** & 0.896** & 0.893** & G.S. length\\
    &Randoms & 0.093 & 0.101 & 0.012* & 0.014* & 0.613** & 0.608** & G.S. length\\
   \cline{2-9}
    &COWTS & 0.080 & 0.054 & 0.004 & 0.002 & 0.690 & 0.655 & G.S. length\\
    &SEMCOWTS & 0.087 & 0.068 & 0.003 & 0.002 & 0.709 & 0.686 & G.S. length\\
    \cline{2-9}
    &TSSuBERT-F & \textbf{0.110**} & \textbf{0.114**} & \textbf{0.017**} & \textbf{0.021**} & 0.693 & 0.676 & G.S. length\\
    &TSSuBERT & 0.099** & 0.111** & 0.013** & 0.018** & \textbf{0.757**} & \textbf{0.741**} & 20 tweets\\
    \midrule
     \midrule
    \multirow{5}{*} {Collection of \cite{DBLP:conf/sigir/RudraGGM018}}  & Randoms & 0.345** & 0.351** & 0.155** & 0.161** & 0.866** & 0.870** & + 200 words per time increment\\
    &COWTS & \textbf{0.455} & \textbf{0.460} & \textbf{0.236} & \textbf{0.242} & 0.929 & 0.931  &+ 200 words per time increment\\
    &SEMCOWTS & 0.453 & 0.456 & 0.230 & 0.234 & 0.928 & 0.930 & + 200 words per time increment\\
   \cline{2-9}
    &TSSuBERT-F & 0.398** & 0.404** & 0.181** & 0.188** & 0.930 & 0.930 & + 200 words per time increment\\
    &TSSuBERT & 0.406* & 0.412* & 0.185** & 0.192** & \textbf{0.935} & \textbf{0.935} & 20 tweets\\
    \bottomrule
\end{tabular}
}
\caption{\label{tab:evaluation_with_sota}\small{Results on the TES 2012-2016 dataset and on the collection in \cite{DBLP:conf/sigir/RudraGGM018}. Results with * are statistically significant different from state of the art (COWTS-SEMCOWTS) (Wilcoxon test\protect\footnotemark, * $p-value$<0.05, ** $p-value$<0.01).  Scores in bold highlight the best model for the considered metric (without considering the Oracle summaries for the TES collection).}}
\end{table*}
\footnotetext{As the distribution was not Gaussian for all models using Shapiro-Wilk test, we decided not to report Student t-test and use Wilcoxon test, as suggested by \cite{DBLP:conf/sigir/UrbanoMM13a}}

\subsubsection{Experimental setup}
\label{subsec:ourCollectionSetup}
First, concerning \textit{the similarity metric used in the salience prediction model},
a preliminary manual evaluation on our dataset suggested that the Cosine score was more appropriate than the ROUGE-2 F-score usually used \cite{DBLP:conf/emnlp/LiuL19,DBLP:conf/aaai/NallapatiZZ17,DBLP:journals/corr/abs-2010-08014}. The objective of our model is thus to predict the cosine scores between input tweets and the Gold Standard. The tweets included in the Oracle summaries are assigned a perfect score during training.

Second, regarding the \textit{pre-trained language model} used for salience prediction, we decided to use the pre-trained DistilBERT \cite{DBLP:journals/corr/abs-1910-01108} model, a lighter version of BERT \cite{DBLP:conf/naacl/DevlinCLT19}. As the BERT model computation is quadratic to the length of the input, we fixed the size limit to 50 tokens to take advantage of the small size of a tweet (more than 99.99\,\% of the tweets in the collection have a length < 50 tokens). 
The \textit{Context} is evaluated for each candidate tweet $c$ as the frequency of each token of the entire tweet vocabulary until $c$ timestamp. We used the BERT trained vocabulary, i.e., 30,522 tokens minus the PAD one. The PAD token is the default token to reach the fixed limit size (50) if the tweet is smaller. In addition, we added 4 tokens for URLs, hashtags, mentions (\symbol{64}), and Retweets. Tweets were pre-processed consequently to replace the four aforementioned instances by the associated token. At last, we thus have 30,525 tokens in the context of each tweet. 

\begin{figure}[!ht]
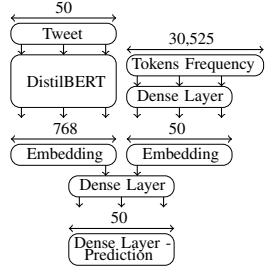

    \centering
    \scalebox{0.7}{
    \small
    \modelArchitecture
    }
    \caption{\label{fig:experiment_architecture}\small{Experimental architecture of the salience part.}}
\end{figure}

Our \textit{neural architecture} is detailed in Figure \ref{fig:experiment_architecture}. 
For each dense layer, we used a ReLu activation function. In addition, as proposed by \cite{DBLP:journals/jmlr/SrivastavaHKSS14}, we added on each of these layers, except the prediction one, a Dropout layer with a probability of 0.5.

To \textit{evaluate} our model, we made a 3-fold cross validation.  The TES 2012-2016 was split in 2 folds containing 9 events, and another one containing 10 events, with approximately the same number of tweets in each fold.
We trained our model with Adam optimizer with parameters $\beta_1$ = 0.9 and $\beta_2$ = 0.999, batch size of 128, a decreasing learning rate as done in \cite{DBLP:conf/emnlp/LiuL19,DBLP:conf/nips/VaswaniSPUJGKP17} with a mean squared error objective function. For each fold, we trained the model for 5 epochs on 90\,\% of the 2 other folds with 10\,\% for validation.

$\lambda_{salience}$ was experimentally set to 0.2.
The $\lambda_{similarity}$ threshold used to avoid redundancy in summaries is set as follows:
\begin{itemize}
    \item $\lambda_{similarity}$ = 0.3 if $length(S_{e,t_0,t})<50$
    \item $\lambda_{similarity} = 0.3* \frac{log(50)}{log(length(S_{e,t_0,t}))}$ else
\end{itemize}
This adaptive threshold aims at reducing the size of the predicted summaries. 

\subsubsection{Results and discussion}
\label{subsec:ourCollectionResults}

The state of the art does not mention approaches that are strictly equivalent to ours and for which it would not be necessary to fix the size of the output summaries. As a consequence, we choose to compare our model with the 2 nearest approaches in terms of tackled issues, i.e., COWTS and SEMCOWTS \cite{DBLP:conf/sigir/RudraGGM018}\footnote{We used the implementation provided by authors on \url{https://github.com/krudra/disaster_summarizer_TWEB_2018}}, and to fix the maximum size of output summaries to the Gold Standard (G.S.) size for the considered time window (see Table~\ref{tab:evaluation_with_sota} Column 9). It should be noted that as opposed to our approach, these two approaches reevaluate the output summaries at each time window and do not propose an incremental approach.
For information, we also report upper and lower bounds on the collection, i.e., our Oracle summaries and a naive baseline defined as the average score for 50 randomly constructed summaries.

For each model, we report ROUGE-1 F, ROUGE-2 F as well as Cos Embed, a semantic similarity measure based on the cosine metric and widely-used embeddings\footnote{We used the word2vec-google-news-300 model provided by gensim \url{https://radimrehurek.com/gensim/models/word2vec.html}.} using Smooth Inverse Frequency (SIF) \cite{DBLP:conf/iclr/AroraLM17} to represent summaries and Gold Standard. Results are presented in Table \ref{tab:evaluation_with_sota}, lines 1 to 7. It should be noted that results are given considering 493 days out of 545. Indeed, 52 days have no associated gold standard (as illustrated in Figure \ref{fig:overview} for time window $t$). Considering these days has no sense in the context where the size of output summaries is fixed.

A first observation is that the TSSuBert model significantly outperforms state-of-the-art approaches on ROUGE metrics when following their requirement to fix the maximum size of the output summaries to the size of the corresponding gold standard (\mbox{TSSuBert-F}, line 6). However this does not reflect, in our opinion, a realistic behaviour in which the size of the gold standard is unknown.
Results on a more realistic protocol are thus reported line 7 (TSSuBert). In this case, we just indicate a maximum number of tweets that can be added at each time window. This limit is introduced for efficiency reasons and fixed to 20 tweets (see Column 8). In this case, the TSSuBert model is able to obtain a summary size close to the gold standard one without knowing it. A second interesting observation is therefore that the normal TSSuBert behaviour still obtains better results than the state-of-the-art approaches.
Moreover, from a semantic point of view (COS Embed scores, columns 7 and 8) the TSSuBert  summaries  are closer to the gold standard than those generated by the semantic-based SEMCOWTS baseline. A third lesson highlighted is that the Randoms method represents a strong baseline, which surprisingly outperforms the two state-of-the art baselines. 
An explanation could be that the terms of the Gold Standard appear in numerous tweets. Results of Oracle summaries show that there is still a large room from improvement. Lastly, it should be noted that experiments made by training the model without the context of tweets lead to very poor results (not reported here), showing thus its fundamental role.

\subsection{Complementary experiments}

\subsubsection{Experimental setup}
To evaluate the robustness of our model, we experimented on the dataset used in \cite{DBLP:conf/sigir/RudraGGM018}, described in  Table~\ref{tab:stats_descs_collections}, line 3. This dataset comprises 11 distinct streams about 3 different events related to natural disasters, for a total of 30 time increments (i.e., 30 days). Gold standard summaries have a fixed size of 200 words for each time increment. It should be noticed that this dataset has the particularity to be composed only of tweets related to information needs of crisis responders, contrary to our more general dataset.
We trained our model on TES 2012-2016, using 26 events out of 28. 2 events were removed since in common with the test collection (i.e., \textit{typhoon hagupit} and \textit{nepal earthquake}).
Parameters are set to the values described in Section \ref{subsec:ourCollectionSetup}, except $\lambda_{salience}$ which is set to 0 (this last value is discussed in the following paragraph).  All tweets with a salience score $>0$ are thus considered in the next step of our model, i.e., the similarity evaluation. 

\subsubsection{Results and discussion}
We report in the second part of Table~\ref{tab:evaluation_with_sota} the results of our model, the COWTS and SEMCOWTS models, and a random baseline built in the same way as in~\ref{subsec:ourCollectionResults}.
The results show, this time, that COWTS and SEMCOWTS baselines are significantly better for ROUGE scores.
The necessity to lower the salience threshold to construct acceptable summaries can give a clue on the lower results obtained with our model. Indeed, the predicted salience scores are lower than for the experiments on our dataset. This means that our model is not adequately trained for the kind of tweets composing the \cite{DBLP:conf/sigir/RudraGGM018} dataset.
However, the results of our model are promising compared with the Randoms baseline and regarding the semantic aspect (i.e., COS Embed measure).

In addition, we evaluated some well-known state-of-the-art models for multi-document summarization such as Textrank \cite{DBLP:conf/emnlp/MihalceaT04}, Lexrank \cite{DBLP:conf/emnlp/ErkanR04}, Centroid \cite{DBLP:journals/ipm/RadevJST04}, and clusterCMRW \cite{DBLP:conf/sigir/WanY08} (not reported due to lack of space). Results are very close to randoms summaries. Even if these  methods are not tweet-oriented, this strengthens our conclusion on the randoms summaries potential to be strong baselines.
\vspace{-6pt}
\section{Conclusion}
\label{sec:conc}

The contribution of this paper is twofold. First, we made publicly available TES 2012-2016, a large evaluation collection for Twitter event summarization that can be used to train neural models. This collection, based on the collection of \cite{DBLP:journals/jasis/Zubiaga18} is composed of 28 events with 3 million tweets on average: it is to our knowledge the largest existing collection for tweet summarization. Second,  we propose a neural model for extractive tweet summarization, TSSuBERT, with several original features:  (i) the use of pre-trained language models as well as tweet context in the form of vocabulary frequencies, (ii) the model automatically fixes the size of output summaries -- this corresponds to a more realistic use case than the one of state-of-the-art models that always output the same size of summaries whether it is useful or not, and (iii) TSSuBERT is an incremental  method that does not require to re-evaluate all the documents at each time increment, leading thus to an efficiency gain.

Preliminary experiments were conducted on two datasets, and showed that our model is able to process huge streams of data as well as to outperform state of the art on a large corpus. Another interesting finding is that random summmaries can be considered as strong baselines for Twitter event summarization and should then be systematically reported in future experiments.

There are several promising research directions for future work. First, we plan to integrate a tweet-specific pre-trained language model such as  BERTweet \cite{DBLP:conf/emnlp/NguyenVN20}. We would also like to learn the thresholds used in the filtering step of our approach, thus avoiding to fix them manually.  Third, we will also explore different ways to integrate tweet context at lower cost in the salience prediction part. Giving the text of tweets selected on preceding time increments to the model is such an avenue to explore. 
Lastly, we considerer this work as a first step of an hybrid approach for abstractive summarization, with the aim of producing more human-readable summaries.

\bibliographystyle{apalike}
\bibliography{biblio}

\begin{thebibliography}{}

\bibitem[Arora et~al., 2017]{DBLP:conf/iclr/AroraLM17}
Arora, S., Liang, Y., and Ma, T. (2017).
\newblock A simple but tough-to-beat baseline for sentence embeddings.
\newblock In {\em 5th International Conference on Learning Representations,
  {ICLR} 2017, Toulon, France, April 24-26, 2017, Conference Track
  Proceedings}. OpenReview.net.

\bibitem[Devlin et~al., 2019]{DBLP:conf/naacl/DevlinCLT19}
Devlin, J., Chang, M., Lee, K., and Toutanova, K. (2019).
\newblock {BERT:} pre-training of deep bidirectional transformers for language
  understanding.
\newblock In Burstein, J., Doran, C., and Solorio, T., editors, {\em
  Proceedings of the 2019 Conference of the North American Chapter of the
  Association for Computational Linguistics: Human Language Technologies,
  {NAACL-HLT} 2019, Minneapolis, MN, USA, June 2-7, 2019, Volume 1 (Long and
  Short Papers)}, pages 4171--4186. Association for Computational Linguistics.

\bibitem[Dou et~al., 2020]{DBLP:journals/corr/abs-2010-08014}
Dou, Z., Liu, P., Hayashi, H., Jiang, Z., and Neubig, G. (2020).
\newblock Gsum: {A} general framework for guided neural abstractive
  summarization.
\newblock {\em CoRR}, abs/2010.08014.

\bibitem[El-Kassas et~al., 2021]{ELKASSAS2021113679}
El-Kassas, W.~S., Salama, C.~R., Rafea, A.~A., and Mohamed, H.~K. (2021).
\newblock Automatic text summarization: A comprehensive survey.
\newblock {\em Expert Systems with Applications}, 165:113679.

\bibitem[Erkan and Radev, 2004]{DBLP:conf/emnlp/ErkanR04}
Erkan, G. and Radev, D.~R. (2004).
\newblock Lexpagerank: Prestige in multi-document text summarization.
\newblock In {\em Proceedings of the 2004 Conference on Empirical Methods in
  Natural Language Processing , {EMNLP} 2004, {A} meeting of SIGDAT, a Special
  Interest Group of the ACL, held in conjunction with {ACL} 2004, 25-26 July
  2004, Barcelona, Spain}, pages 365--371. {ACL}.

\bibitem[Ghalandari et~al., 2020]{DBLP:conf/acl/GhalandariHPGI20}
Ghalandari, D.~G., Hokamp, C., Pham, N.~T., Glover, J., and Ifrim, G. (2020).
\newblock A large-scale multi-document summarization dataset from the wikipedia
  current events portal.
\newblock In Jurafsky, D., Chai, J., Schluter, N., and Tetreault, J.~R.,
  editors, {\em Proceedings of the 58th Annual Meeting of the Association for
  Computational Linguistics, {ACL} 2020, Online, July 5-10, 2020}, pages
  1302--1308. Association for Computational Linguistics.

\bibitem[Graff et~al., 2003]{graff2003english}
Graff, D., Kong, J., Chen, K., and Maeda, K. (2003).
\newblock English gigaword.
\newblock {\em Linguistic Data Consortium, Philadelphia}, 4(1):34.

\bibitem[Hermann et~al., 2015]{Hermann:2015}
Hermann, K.~M., Kocisky, T., Grefenstette, E., Espeholt, L., Kay, W., Suleyman,
  M., and Blunsom, P. (2015).
\newblock Teaching machines to read and comprehend.
\newblock In {\em Advances in neural information processing systems}, pages
  1693--1701.

\bibitem[Hubert et~al., 2017]{hubert:hal-02319737}
Hubert, G., Moreno, J.~G., Pinel-Sauvagnat, K., and Pitarch, Y. (2017).
\newblock {Some thoughts from IRIT about the scenario A of the TREC RTS 2016
  and 2017 tracks}.
\newblock In {\em {26th Text REtrieval Conference (TREC 2017)}}, pages 1--12,
  Gaithersburg, Maryland, United States.

\bibitem[Li and Zhang, 2020]{DBLP:conf/www/LiZ20}
Li, Q. and Zhang, Q. (2020).
\newblock Abstractive event summarization on twitter.
\newblock In Seghrouchni, A. E.~F., Sukthankar, G., Liu, T., and van Steen, M.,
  editors, {\em Companion of The 2020 Web Conference 2020, Taipei, Taiwan,
  April 20-24, 2020}, pages 22--23. {ACM} / {IW3C2}.

\bibitem[Li and Zhang, 2021]{li2021twitter}
Li, Q. and Zhang, Q. (2021).
\newblock Twitter event summarization by exploiting semantic terms and graph
  network.
\newblock In {\em Proceedings of the The Thirty-Third Annual Conference on
  Innovative Applications of Artificial Intelligence (IAAI-21)}.

\bibitem[Lin et~al., 2014]{lin2014overview}
Lin, J., Efron, M., Wang, Y., and Sherman, G. (2014).
\newblock Overview of the trec-2014 microblog track.
\newblock Technical report, MARYLAND UNIV COLLEGE PARK.

\bibitem[Liu and Lapata, 2019]{DBLP:conf/emnlp/LiuL19}
Liu, Y. and Lapata, M. (2019).
\newblock Text summarization with pretrained encoders.
\newblock In Inui, K., Jiang, J., Ng, V., and Wan, X., editors, {\em
  Proceedings of the 2019 Conference on Empirical Methods in Natural Language
  Processing and the 9th International Joint Conference on Natural Language
  Processing, {EMNLP-IJCNLP} 2019, Hong Kong, China, November 3-7, 2019}, pages
  3728--3738. Association for Computational Linguistics.

\bibitem[McCreadie et~al., 2020]{mccreadie2020incident}
McCreadie, R., Buntain, C., and Soboroff, I. (2020).
\newblock Incident streams 2019: Actionable insights and how to find them.
\newblock In {\em Proceedings of the International ISCRAM Conference}.

\bibitem[Mihalcea and Tarau, 2004]{DBLP:conf/emnlp/MihalceaT04}
Mihalcea, R. and Tarau, P. (2004).
\newblock Textrank: Bringing order into text.
\newblock In {\em Proceedings of the 2004 Conference on Empirical Methods in
  Natural Language Processing , {EMNLP} 2004, {A} meeting of SIGDAT, a Special
  Interest Group of the ACL, held in conjunction with {ACL} 2004, 25-26 July
  2004, Barcelona, Spain}, pages 404--411. {ACL}.

\bibitem[Nallapati et~al., 2017]{DBLP:conf/aaai/NallapatiZZ17}
Nallapati, R., Zhai, F., and Zhou, B. (2017).
\newblock Summarunner: {A} recurrent neural network based sequence model for
  extractive summarization of documents.
\newblock In Singh, S.~P. and Markovitch, S., editors, {\em Proceedings of the
  Thirty-First {AAAI} Conference on Artificial Intelligence, February 4-9,
  2017, San Francisco, California, {USA}}, pages 3075--3081. {AAAI} Press.

\bibitem[Narayan et~al., 2018]{DBLP:conf/emnlp/NarayanCL18}
Narayan, S., Cohen, S.~B., and Lapata, M. (2018).
\newblock Don't give me the details, just the summary! topic-aware
  convolutional neural networks for extreme summarization.
\newblock In Riloff, E., Chiang, D., Hockenmaier, J., and Tsujii, J., editors,
  {\em Proceedings of the 2018 Conference on Empirical Methods in Natural
  Language Processing, Brussels, Belgium, October 31 - November 4, 2018}, pages
  1797--1807. Association for Computational Linguistics.

\bibitem[Nguyen et~al., 2020]{DBLP:conf/emnlp/NguyenVN20}
Nguyen, D.~Q., Vu, T., and Nguyen, A.~T. (2020).
\newblock Bertweet: {A} pre-trained language model for english tweets.
\newblock In Liu, Q. and Schlangen, D., editors, {\em Proceedings of the 2020
  Conference on Empirical Methods in Natural Language Processing: System
  Demonstrations, {EMNLP} 2020 - Demos, Online, November 16-20, 2020}, pages
  9--14. Association for Computational Linguistics.

\bibitem[Olariu, 2014]{DBLP:conf/eacl/Olariu14}
Olariu, A. (2014).
\newblock Efficient online summarization of microblogging streams.
\newblock In Bouma, G. and Parmentier, Y., editors, {\em Proceedings of the
  14th Conference of the European Chapter of the Association for Computational
  Linguistics, {EACL} 2014, April 26-30, 2014, Gothenburg, Sweden}, pages
  236--240. The Association for Computer Linguistics.

\bibitem[Qi et~al., 2020]{DBLP:conf/emnlp/QiYGLDCZ020}
Qi, W., Yan, Y., Gong, Y., Liu, D., Duan, N., Chen, J., Zhang, R., and Zhou, M.
  (2020).
\newblock Prophetnet: Predicting future n-gram for sequence-to-sequence
  pre-training.
\newblock In Cohn, T., He, Y., and Liu, Y., editors, {\em Proceedings of the
  2020 Conference on Empirical Methods in Natural Language Processing:
  Findings, {EMNLP} 2020, Online Event, 16-20 November 2020}, pages 2401--2410.
  Association for Computational Linguistics.

\bibitem[Radev et~al., 2004]{DBLP:journals/ipm/RadevJST04}
Radev, D.~R., Jing, H., Sty, M., and Tam, D. (2004).
\newblock Centroid-based summarization of multiple documents.
\newblock {\em Inf. Process. Manag.}, 40(6):919--938.

\bibitem[Rudra et~al., 2018a]{DBLP:journals/tweb/RudraGGG18}
Rudra, K., Ganguly, N., Goyal, P., and Ghosh, S. (2018a).
\newblock Extracting and summarizing situational information from the twitter
  social media during disasters.
\newblock {\em {ACM} Trans. Web}, 12(3):17:1--17:35.

\bibitem[Rudra et~al., 2015]{DBLP:conf/cikm/RudraGGGG15}
Rudra, K., Ghosh, S., Ganguly, N., Goyal, P., and Ghosh, S. (2015).
\newblock Extracting situational information from microblogs during disaster
  events: a classification-summarization approach.
\newblock In Bailey, J., Moffat, A., Aggarwal, C.~C., de~Rijke, M., Kumar, R.,
  Murdock, V., Sellis, T.~K., and Yu, J.~X., editors, {\em Proceedings of the
  24th {ACM} International Conference on Information and Knowledge Management,
  {CIKM} 2015, Melbourne, VIC, Australia, October 19 - 23, 2015}, pages
  583--592. {ACM}.

\bibitem[Rudra et~al., 2019]{DBLP:journals/tcss/RudraGGIM19}
Rudra, K., Goyal, P., Ganguly, N., Imran, M., and Mitra, P. (2019).
\newblock Summarizing situational tweets in crisis scenarios: An
  extractive-abstractive approach.
\newblock {\em {IEEE} Trans. Comput. Soc. Syst.}, 6(5):981--993.

\bibitem[Rudra et~al., 2018b]{DBLP:conf/sigir/RudraGGM018}
Rudra, K., Goyal, P., Ganguly, N., Mitra, P., and Imran, M. (2018b).
\newblock Identifying sub-events and summarizing disaster-related information
  from microblogs.
\newblock In Collins{-}Thompson, K., Mei, Q., Davison, B.~D., Liu, Y., and
  Yilmaz, E., editors, {\em The 41st International {ACM} {SIGIR} Conference on
  Research {\&} Development in Information Retrieval, {SIGIR} 2018, Ann Arbor,
  MI, USA, July 08-12, 2018}, pages 265--274. {ACM}.

\bibitem[Rudrapal et~al., 2018]{DBLP:journals/jips/RudrapalDB18}
Rudrapal, D., Das, A., and Bhattacharya, B. (2018).
\newblock A survey on automatic twitter event summarization.
\newblock {\em J. Inf. Process. Syst.}, 14(1):79--100.

\bibitem[Rush et~al., 2015]{Rush_2015}
Rush, A.~M., Chopra, S., and Weston, J. (2015).
\newblock A neural attention model for abstractive sentence summarization.
\newblock {\em Proceedings of the 2015 Conference on Empirical Methods in
  Natural Language Processing}.

\bibitem[Sanh et~al., 2019]{DBLP:journals/corr/abs-1910-01108}
Sanh, V., Debut, L., Chaumond, J., and Wolf, T. (2019).
\newblock Distilbert, a distilled version of {BERT:} smaller, faster, cheaper
  and lighter.
\newblock {\em CoRR}, abs/1910.01108.

\bibitem[See et~al., 2017]{DBLP:conf/acl/SeeLM17}
See, A., Liu, P.~J., and Manning, C.~D. (2017).
\newblock Get to the point: Summarization with pointer-generator networks.
\newblock In Barzilay, R. and Kan, M., editors, {\em Proceedings of the 55th
  Annual Meeting of the Association for Computational Linguistics, {ACL} 2017,
  Vancouver, Canada, July 30 - August 4, Volume 1: Long Papers}, pages
  1073--1083. Association for Computational Linguistics.

\bibitem[Sequiera et~al., 2018]{DBLP:conf/trec/SequieraTL18}
Sequiera, R., Tan, L., and Lin, J. (2018).
\newblock Overview of the {TREC} 2018 real-time summarization track.
\newblock In Voorhees, E.~M. and Ellis, A., editors, {\em Proceedings of the
  Twenty-Seventh Text REtrieval Conference, {TREC} 2018, Gaithersburg,
  Maryland, USA, November 14-16, 2018}, volume 500-331 of {\em {NIST} Special
  Publication}. National Institute of Standards and Technology {(NIST)}.

\bibitem[Song et~al., 2020]{DBLP:conf/aaai/SongWFL020}
Song, K., Wang, B., Feng, Z., Liu, R., and Liu, F. (2020).
\newblock Controlling the amount of verbatim copying in abstractive
  summarization.
\newblock In {\em The Thirty-Fourth {AAAI} Conference on Artificial
  Intelligence, {AAAI} 2020, The Thirty-Second Innovative Applications of
  Artificial Intelligence Conference, {IAAI} 2020, The Tenth {AAAI} Symposium
  on Educational Advances in Artificial Intelligence, {EAAI} 2020, New York,
  NY, USA, February 7-12, 2020}, pages 8902--8909. {AAAI} Press.

\bibitem[Srivastava et~al., 2014]{DBLP:journals/jmlr/SrivastavaHKSS14}
Srivastava, N., Hinton, G.~E., Krizhevsky, A., Sutskever, I., and
  Salakhutdinov, R. (2014).
\newblock Dropout: a simple way to prevent neural networks from overfitting.
\newblock {\em J. Mach. Learn. Res.}, 15(1):1929--1958.

\bibitem[Urbano et~al., 2013]{DBLP:conf/sigir/UrbanoMM13a}
Urbano, J., Marrero, M., and Mart{\'{\i}}n, D. (2013).
\newblock A comparison of the optimality of statistical significance tests for
  information retrieval evaluation.
\newblock In Jones, G. J.~F., Sheridan, P., Kelly, D., de~Rijke, M., and Sakai,
  T., editors, {\em The 36th International {ACM} {SIGIR} conference on research
  and development in Information Retrieval, {SIGIR} '13, Dublin, Ireland - July
  28 - August 01, 2013}, pages 925--928. {ACM}.

\bibitem[Vaswani et~al., 2017]{DBLP:conf/nips/VaswaniSPUJGKP17}
Vaswani, A., Shazeer, N., Parmar, N., Uszkoreit, J., Jones, L., Gomez, A.~N.,
  Kaiser, L., and Polosukhin, I. (2017).
\newblock Attention is all you need.
\newblock In Guyon, I., von Luxburg, U., Bengio, S., Wallach, H.~M., Fergus,
  R., Vishwanathan, S. V.~N., and Garnett, R., editors, {\em Advances in Neural
  Information Processing Systems 30: Annual Conference on Neural Information
  Processing Systems 2017, December 4-9, 2017, Long Beach, CA, {USA}}, pages
  5998--6008.

\bibitem[Wan and Yang, 2008]{DBLP:conf/sigir/WanY08}
Wan, X. and Yang, J. (2008).
\newblock Multi-document summarization using cluster-based link analysis.
\newblock In Myaeng, S., Oard, D.~W., Sebastiani, F., Chua, T., and Leong, M.,
  editors, {\em Proceedings of the 31st Annual International {ACM} {SIGIR}
  Conference on Research and Development in Information Retrieval, {SIGIR}
  2008, Singapore, July 20-24, 2008}, pages 299--306. {ACM}.

\bibitem[Zhang et~al., 2020]{DBLP:conf/icml/ZhangZSL20}
Zhang, J., Zhao, Y., Saleh, M., and Liu, P.~J. (2020).
\newblock {PEGASUS:} pre-training with extracted gap-sentences for abstractive
  summarization.
\newblock In {\em Proceedings of the 37th International Conference on Machine
  Learning, {ICML} 2020, 13-18 July 2020, Virtual Event}, volume 119 of {\em
  Proceedings of Machine Learning Research}, pages 11328--11339. {PMLR}.

\bibitem[Zhang et~al., 2013]{DBLP:journals/taslp/ZhangLGY13}
Zhang, R., Li, W., Gao, D., and You, O. (2013).
\newblock Automatic twitter topic summarization with speech acts.
\newblock {\em {IEEE} Trans. Speech Audio Process.}, 21(3):649--658.

\bibitem[Zhong et~al., 2020]{DBLP:conf/acl/ZhongLCWQH20}
Zhong, M., Liu, P., Chen, Y., Wang, D., Qiu, X., and Huang, X. (2020).
\newblock Extractive summarization as text matching.
\newblock In Jurafsky, D., Chai, J., Schluter, N., and Tetreault, J.~R.,
  editors, {\em Proceedings of the 58th Annual Meeting of the Association for
  Computational Linguistics, {ACL} 2020, Online, July 5-10, 2020}, pages
  6197--6208. Association for Computational Linguistics.

\bibitem[Zubiaga, 2018]{DBLP:journals/jasis/Zubiaga18}
Zubiaga, A. (2018).
\newblock A longitudinal assessment of the persistence of twitter datasets.
\newblock {\em J. Assoc. Inf. Sci. Technol.}, 69(8):974--984.

\end{thebibliography}

\end{document}